\documentclass[conference]{IEEEtran}
\IEEEoverridecommandlockouts

\usepackage{cite}
\usepackage{amsmath,amssymb,amsfonts}
\usepackage{algorithmic}
\usepackage{graphicx}
\usepackage{textcomp}
\usepackage{colortbl} 
\usepackage{verbatim}
\usepackage{graphicx} 
\usepackage{threeparttable}

\usepackage{array}        
\usepackage{booktabs} 
\usepackage{ragged2e} 
\usepackage{makecell} 
\usepackage{colortbl} 
\usepackage{graphicx}
\usepackage{subcaption}
\usepackage{float}
\usepackage[table,xcdraw]{xcolor} 
\usepackage{xcolor}
\usepackage{xspace} 
\def\BibTeX{{\rm B\kern-.05em{\sc i\kern-.025em b}\kern-.08em
    T\kern-.1667em\lower.7ex\hbox{E}\kern-.125emX}}

\begin{document}

\title{SecureInfer: Heterogeneous TEE-GPU Architecture for Privacy-Critical Tensors for Large Language Model Deployment\\
}


\author{
\IEEEauthorblockN{Tushar Nayan}
\IEEEauthorblockA{\textit{Florida International University}\\
Miami, FL, USA\\
tnaya002@fiu.edu\\
ORCID: 0009-0003-4317-2371}
\and
\IEEEauthorblockN{Ziqi Zhang}
\IEEEauthorblockA{\textit{University of Illinois Urbana-Champaign}\\
Urbana, IL, USA\\
ziqi24@illinois.edu\\
ORCID: 0000-0001-8493-0261} 
\and
\IEEEauthorblockN{Ruimin Sun}
\IEEEauthorblockA{\textit{Florida International University}\\
Miami, FL, USA\\
rsun@fiu.edu\\
ORCID: 0000-0003-2940-5549}
}

\maketitle
\pagestyle{plain}

\begin{abstract}
With the increasing deployment of Large Language Models (LLMs) on mobile and edge platforms, securing them against model extraction attacks has become a pressing concern. However, protecting model privacy without sacrificing the performance benefits of untrusted AI accelerators, such as GPUs, presents a challenging trade-off. In this paper, we initiate the study of high-performance execution on LLMs and present \textbf{SecureInfer}, a hybrid framework that leverages a heterogeneous Trusted Execution Environments (TEEs)-GPU architecture to isolate privacy-critical components while offloading compute-intensive operations to untrusted accelerators. Building upon an outsourcing scheme, SecureInfer adopts an information-theoretic and threat-informed partitioning strategy: security-sensitive components, including non-linear layers, projection of attention head, FNN transformations, and LoRA adapters are executed inside an SGX enclave, while other linear operations  (matrix multiplication) are performed on the GPU after encryption and are securely restored within the enclave. We implement a prototype of SecureInfer using the LLaMA-2 model and evaluate it across performance and security metrics. Our results show that SecureInfer offers strong security guarantees with reasonable performance, offering a practical solution for secure on-device model inference.
\end{abstract}

\section{Introduction} \label{sec:intro}

Large Language Models (LLMs) have become foundational to tasks such as voice assistance, summarization, and intelligent edge computing. While LLMs were traditionally confined to cloud deployments, advances in edge AI hardware, such as NVIDIA Jetson Orin, Qualcomm AI Engine, and Apple Neural Engine, now enable real-time execution of quantized models (e.g., INT4/INT8) on mobile and embedded devices. This allows responsive, private inference for models like LLaMA-2 7B, Distilled GPT-2, and TinyLLM variants~\cite{zhang2024tinyllama}, with generation speeds reaching 5–20 tokens per second.

However, deploying LLMs directly on-device introduces significant model extraction risks. Adversaries with local access can exploit various stages of the inference pipeline to reconstruct proprietary models. At the \textbf{block level}, deeper layer representations leak semantic content~\cite{teeSlice}, at the \textbf{tensor level}, unprotected attention scores or KV caches reveal focus patterns and internal dynamics~\cite{lord, teeSlice}, and at the \textbf{output level}, full logits expose rich statistical distributions that can be exploited for surrogate model training~\cite{Knockoff, ML-Doctor}.

Existing defenses against model extraction either assume remote-only access (e.g., API rate limiting, output obfuscation) or apply heuristic layer isolation strategies originally designed for convolutional neural networks (CNNs) and deep neural networks (DNNs)~\cite{DarkneTZ, shadownet}. These approaches fall short in the context of transformer-based LLMs, where high-dimensional representations, long-range attention, and fine-tuning mechanisms such as LoRA present new and subtle leakage surfaces.

To address these challenges, we present \textbf{SecureInfer}, a hybrid Trusted Execution Environments (TEEs)–GPU co-execution framework that combines hardware-backed isolation with performance-aware model partitioning. SecureInfer uses a \textbf{threat-informed, information-theoretic }partitioning strategy to isolate sensitive components, such as attention projections, feed-forward blocks, and LoRA adapters, within an Intel SGX enclave, while securely offloading compute-intensive matrix operations to the GPU. This design preserves privacy without incurring prohibitive latency or energy costs.

We prototype SecureInfer on a customized LLaMA-2 model and evaluate its performance, energy efficiency, and extraction resistance. Compared to full TEE-based inference, SecureInfer achieves up to \textbf{4.7× speedup }while significantly reducing extraction fidelity, with \textbf{BLEU score degradation under 0.12} in black-box attack scenarios.

\medskip
\noindent \textbf{Our contributions are:}

\begin{itemize}
    \item We present a secure LLM inference framework that combines enclave isolation with GPU acceleration.
    \item We propose a fine-grained, threat-aware partitioning method tailored to transformer models.
    \item We demonstrate the effectiveness of SecureInfer on LLaMA-2, showing strong protection with modest overhead.
\end{itemize}

\section{Background and Related Work}

\textbf{LLMs on Edge Devices.} Transformer-based LLMs such as GPT, LLaMA, and Falcon~\cite{gpt2,llama,falcon} achieve strong performance on tasks like QA, summarization, and dialogue. Optimized variants (e.g., LLaMA-2 in GGUF format) now enable real-time inference (5–20 tokens/s) on edge devices like NVIDIA Jetson Orin. However, deploying LLMs on untrusted platforms exposes them to model extraction attacks~\cite{Knockoff,ML-Doctor,teeSlice, lord, zheng2024inputsnatchstealinginputllm} where adversaries replicate functionality via query access. Balancing security with limited compute and energy remains a key challenge.

\textbf{Model Extraction Attacks on On-Device LLMs.}
In these attacks, adversaries can recover model parameters or approximate a deployed model by observing its outputs, side channels, or internal states during inference. Prior work has demonstrated that even black-box access to LLMs can leak architectural details, hyperparameters, and even full weights~\cite{dong2025depthgivesfalsesense, TMI, wang2025pradablackboxllmadaptation}. Adversaries can also have access to the execution environment, making grey-box attacks potent. For instance, attackers can monitor intermediate activations or kernel-level GPU traffic to infer sensitive model components. In transformer-based models, projections such as Q/K/V matrices and LoRA adapters are especially vulnerable due to their structural uniqueness and linear separability.

\textbf{Limitations of TEEs.} 
TEEs such as Intel SGX and ARM TrustZone provide isolated execution environments that ensure confidentiality and integrity even against privileged software adversaries~\cite{armtrustzone}. Secure model execution inside TEEs~\cite{armtrustzone} has emerged as a promising alternative, but fully enclosing large models like LLMs within a TEE is often infeasible due to memory and performance constraints.

Prior work, such as DarkneTZ~\cite{DarkneTZ}, ShadowNet~\cite{shadownet}, SOTER~\cite{SOTER}, SALOM~\cite{slalom}, and Serdab~\cite{serdab}, primarily focused on CNNs and traditional DNNs. These methods typically rely on coarse-grained heuristics, such as layer depth, weight magnitude, or early-exit layers, to determine which parts of the model to isolate within a TEE. However, unlike CNNs, where computation and semantics tend to be hierarchically structured, LLMs exhibit distributed semantic encoding across multiple layers, attention heads, and even token positions. In LLMs, security-sensitive information such as learned embeddings, attention key-value dynamics, and fine-tuned adaptation parameters is often interleaved throughout the architecture. Moreover, attention-based models are characterized by tight inter-layer dependencies, making it difficult to isolate individual layers without impacting functional integrity or incurring significant communication overhead. Consequently, direct adaptation of CNN-style partitioning to LLMs leads to ineffective protection, unnecessary latency, and memory bottlenecks.

\section{Threat Model}
We consider an adversary aiming to extract confidential model parameters or reproduce a surrogate model by exploiting the on-device inference process. Our model reflects realistic deployment scenarios where LLMs are executed locally on edge or consumer-grade devices, such as smartphones, IoT hubs, or embedded AI modules.

\textbf{Adversary Capabilities.}
The adversary has \textbf{user-level access} to the device, with the ability to interact with the deployed model through a public API (black-box access), 
and collect and analyze final model outputs (logits, tokens) for statistical reconstruction. 
We assume the attacker \textbf{cannot directly compromise the TEE} (e.g., SGX), but may attempt to infer sensitive information via repeated inference queries to induce and analyze internal state changes (grey-box access).

\textbf{Assets.} The primary asset is the deployed LLM, including:
full model weights and fine-tuned parameters, 
adaptation modules such as LoRA layers, 
attention activations and intermediate token embeddings. 

SecureInfer aims to protect against adversaries attempting to reconstruct model parameters or weight matrices, train a high-fidelity surrogate model via black-box queries, or 
extract LoRA or adaptation-specific deltas from memory. To achieve this, SecureInfer enforces confidentiality of security-sensitive submodules via TEE isolation and limits information leakage through secure GPU offloading and output perturbation techniques.

\section{SecureInfer's Design} \label{se:design}

\subsection{Design Goals and Partitioning Strategy}
SecureInfer adopts a \textit{threat-informed and information-theoretic partitioning strategy} tailored to the structure and sensitivity of LLMs. Unlike prior works that use static depth-based rules, our approach partitions model components based on: \textbf{(i) sensitivity}, defined as the susceptibility of components to model extraction attacks, guided by prior threat models~\cite{llmSurvey,Knockoff,ML-Doctor,teeSlice,YAO2024100211} and empirical studies~\cite{YAO2024100211,llmSurvey}; \textbf{(ii) computational intensity}, reflecting the execution cost incurred by edge devices. 
Our dual-objective design enables fine-grained, deployable partitions that balance security and performance across heterogeneous platforms.

Table~\ref{tab:layer_partition_domains} summarizes our component-level mapping across SGX, GPU, and CPU domains. Operations that expose intermediate representations or are vulnerable to adversarial analysis, such as Q/K/V projections, attention score computations, FFN blocks, and LoRA adapters, are executed within the SGX enclave. Meanwhile, performance-sensitive yet predictable components, such as activation functions (e.g., SiLU), residual connections, and positional encodings, are offloaded to the GPU. Input/output handling is delegated to the CPU, which does not interact with sensitive model parameters.

\subsection{Memory-and Latency-Aware Design Considerations}
Edge deployments of LLMs are constrained by limited memory (especially in SGX enclaves) and tight latency budgets. SecureInfer incorporates several design features to address these challenges.

\textbf{Chunked Secure Tensor Execution}: SGX buffer sizes are bounded by \texttt{STORE\_CHUNK\_ELEM}, limiting tensor element counts per enclave transfer. For long sequences, the required chunk size scales with 
\texttt{[batch × seq\_len × hidden\_dim]}, necessitating careful memory budgeting to avoid EPC paging overhead.

\textbf{Lightweight Encryption for Untrusted Memory}: For performance-critical components that must reside outside the enclave (e.g., embedding and logits projection layers), we apply XOR-based encryption. These tensors are decrypted just-in-time within the enclave and securely zeroed afterward.

\textbf{One-Time Pad Masking for Cross-Domain Transfers}: Inspired by TEESlice~\cite{teeSlice}, we use a one-time pad masking mechanism to protect intermediate values transferred across GPU, CPU, and SGX with minimal communication overhead.

These memory-aware adaptations are essential to maintaining real-time inference performance and TEE feasibility on resource-constrained edge devices.

\begin{table}[t]
\centering
\small
\caption{Model Components Placement Across Execution Domains}
\label{tab:layer_partition_domains}
\begin{tabular}{@{}lccc@{}}
\toprule
\textbf{Layer Type} & \textbf{SGX} & \textbf{GPU} & \textbf{CPU} \\
\midrule
\texttt{EmbeddingLayer}      &             & \checkmark   &             \\
\texttt{Linear (Q/K/V)}      & \checkmark  &              &             \\
\texttt{Attention Multiply}  & \checkmark  &              &             \\
\texttt{Linear (FFN)}        & \checkmark  &              &             \\
\texttt{LoRA Adapters}       & \checkmark  &              &             \\
\texttt{SiLU/ReLU}           &             & \checkmark   &             \\
\texttt{Add/Multiply}        &             & \checkmark   &             \\
\texttt{IdentityLayer}       &             & \checkmark   &             \\
\texttt{Input/Output Layer}  &             &              & \checkmark  \\
\bottomrule
\end{tabular}
\end{table}

\subsection{Model Component Placement Strategy}
SecureInfer partitions LLaMA model components across execution domains according to their role and sensitivity, as shown in Table~\ref{tab:layer_partition_domains}. Sensitive operations are contained within the SGX enclave, performance-optimized layers are offloaded to the GPU, and non-critical control flows are assigned to the CPU.

This hybrid placement minimizes enclave workload while securing critical tensors, striking a balance between privacy, efficiency, and deployment feasibility. The design also accommodates future extensions such as adaptive runtime scheduling and energy-aware inference optimization.

\section{Experimental Evaluation} 
\label{se:Experiments}


\begin{figure*}[t]
  \centering
  \begin{subfigure}[b]{0.32\textwidth}
    \includegraphics[width=\textwidth]{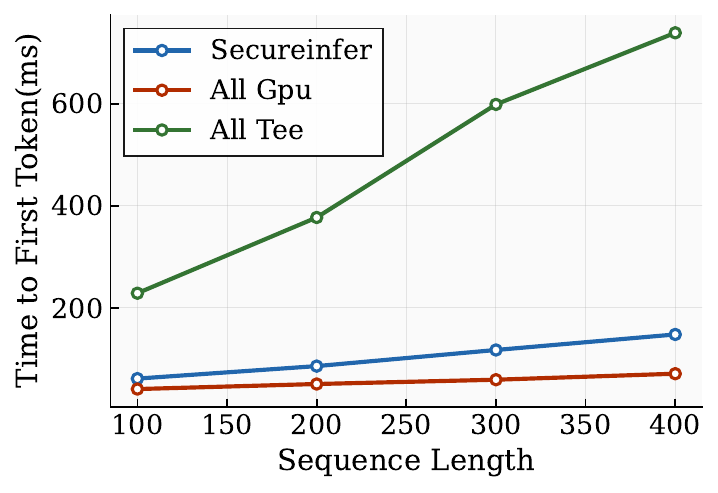}
    \caption{TTFTvs.Seq. Len}
  \end{subfigure}
  \hfill
  \begin{subfigure}[b]{0.32\textwidth}
    \includegraphics[width=\textwidth]{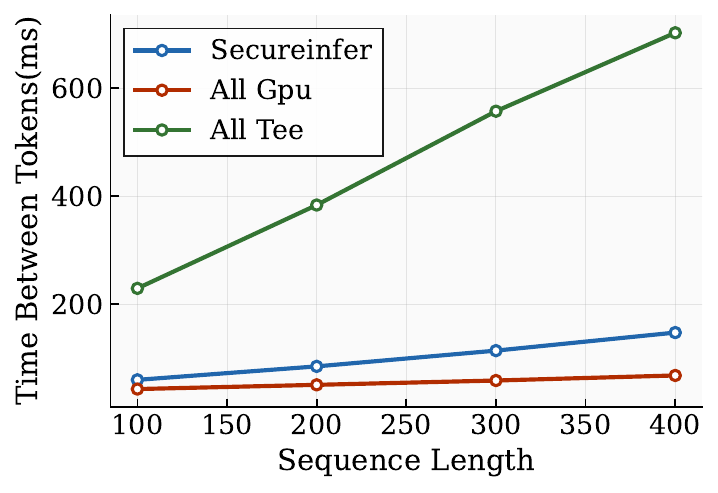}
    \caption{TBTvs.Seq. Len}
  \end{subfigure}
  \hfill
  \begin{subfigure}[b]{0.32\textwidth}
    \includegraphics[width=\textwidth]{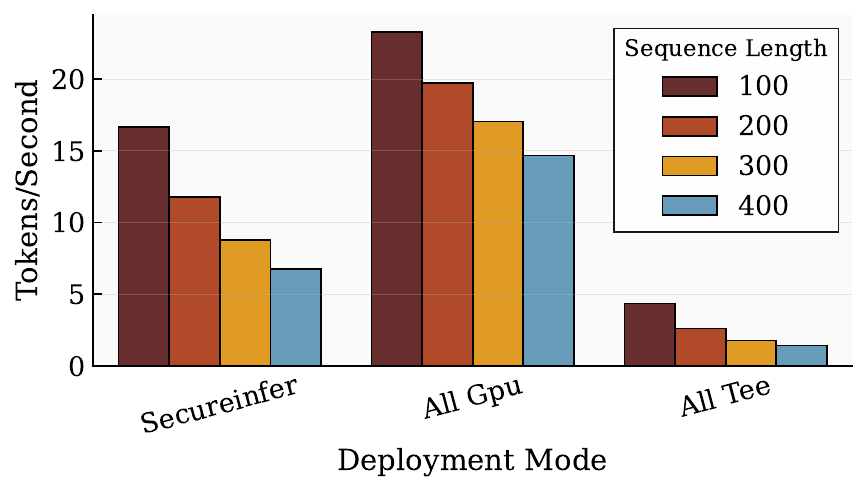}
    \caption{Tokens/sec}
  \end{subfigure}

  \vspace{0.2cm}

  \begin{subfigure}[b]{0.32\textwidth}
    \includegraphics[width=\textwidth]{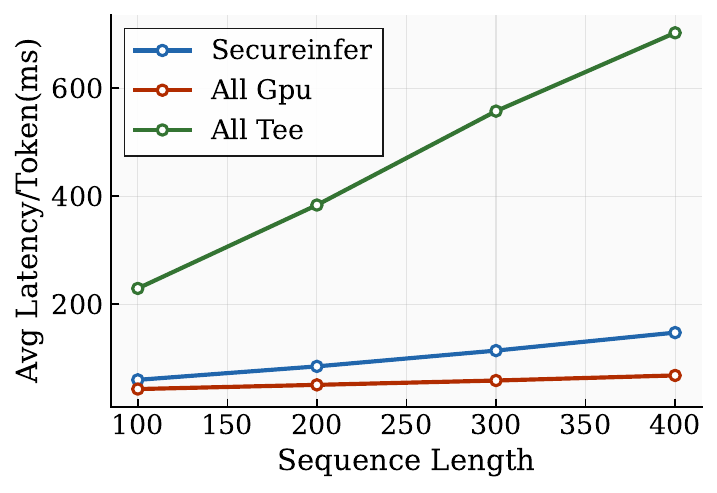}
    \caption{Avg.Latency/Token}
  \end{subfigure}
  \hfill
  \begin{subfigure}[b]{0.32\textwidth}
    \includegraphics[width=\textwidth]{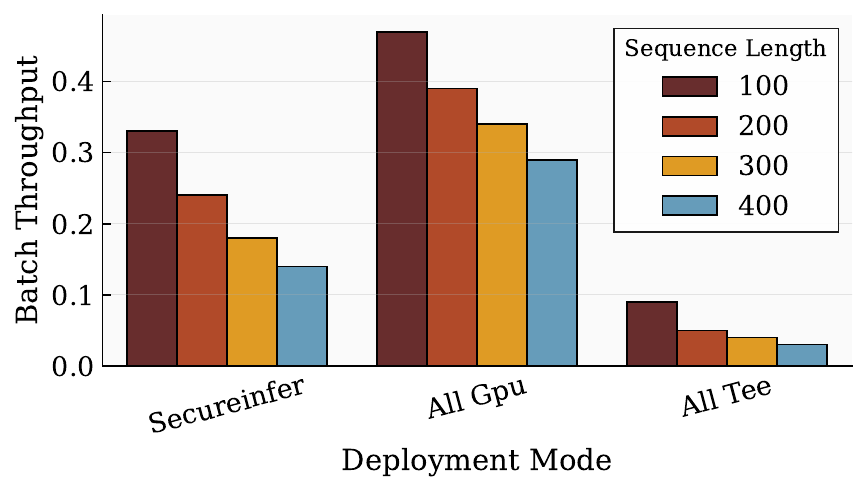}
    \caption{Batch Throughput}
  \end{subfigure}
  \hfill
  \begin{subfigure}[b]{0.32\textwidth}
    \includegraphics[width=\textwidth]{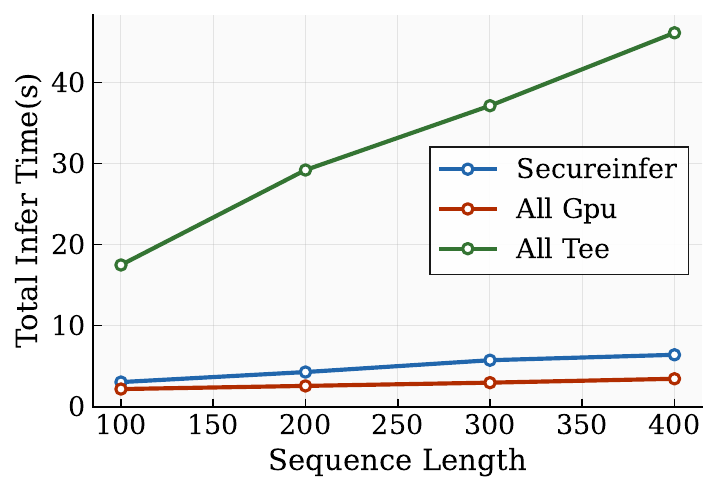}
    \caption{Total InferTime}
  \end{subfigure}

  \caption{Comparison of latency and throughput metrics across deployment modes (\textit{All GPU}, \textit{All TEE}, \textit{SecureInfer}) varying sequence lengths.}
  \label{fig:latency_throughput_comparison}
\end{figure*}



We evaluate SecureInfer across performance, energy, and security dimensions. Our study compares SecureInfer against GPU-only, TEE-only, and CPU-only baselines to answer the following \textbf{research questions}: 
(i) How does SecureInfer’s latency and throughput scale with input size? 
(ii) What is the energy cost of hybrid execution? 
(iii) How does SecureInfer defend against model extraction attacks? 
and (iv) Does SecureInfer preserve model accuracy and behavior? 



\subsection{Experimental Setup and Model Configuration}  
\label{se:Configurations}

We conduct controlled inference experiments on a locally hosted platform equipped with an Intel Core i9-14900KF CPU, an NVIDIA RTX 4090 GPU (24 GB VRAM), and Intel SGX (SGX2, Ubuntu 22.04.5). All experiments use a customized LLaMA-2 model~\cite{llama} with a fixed prompt, an input sequence length of 300, and a batch size of 1. The model uses 32 attention heads and a four-layer decoder-only transformer. Output generation is capped at 50 tokens. Each configuration is executed ten times to average runtime variability.

To understand the impact of deployment parameters on edge feasibility, we evaluate the effects of model size, quantization, batch size, parallelism, and frequency settings. Table~\ref{tab:impact_parameters} summarizes their impact on performance, temperature, power, and output quality.

\begin{table}[htbp]
    \centering
    \caption{Impact of different parameters on the performance, temperature, power, and quality of an LLM inference on edge systems}
    \label{tab:impact_parameters}
    \begin{tabular}{lcccc}
        \toprule
        \textbf{Configuration parameters} & \textbf{Perf} & \textbf{Temp} & \textbf{Power} & \textbf{Quality} \\
        \midrule
        Model Size (e.g., 70B$\to$7B) & $\uparrow$ & $\downarrow$ & $\downarrow$ & $\downarrow$ \\
        Quantization (e.g., FP32$\to$FP16) & $\uparrow$ & $\downarrow$ & $\downarrow$ & $\downarrow$ \\
        Batch Size (e.g., 32$\to$16$\to$4) & $\downarrow$ & $\downarrow$ & $\downarrow$ & -- \\
        Parallelism (e.g., TP4$\to$TP1) & $\downarrow$ & $\uparrow$ & $\downarrow$ & -- \\
        Frequency (e.g., 2GHz$\to$1GHz) & $\downarrow$ & $\downarrow$ & $\downarrow$ & -- \\
        \bottomrule
    \end{tabular}
\end{table}


Prompt-phase latency profiling reveals that Layer 0 operations—especially key/value cache initialization—dominate runtime, contributing 54.9\% of the total prompt latency, while embedding layers account for 27.6\% (Table~\ref{tab:latency_breakdown}). This imbalance is more pronounced under higher precision settings, underscoring the need to optimize early-token operations in hybrid deployments.

\begin{table}[ht]
\centering
\caption{Latency breakdown (ms) of core operations during the prompt phase.}
\label{tab:latency_breakdown}
\begin{tabular}{l c}
\toprule
\textbf{Operation} & \textbf{Latency (ms)} \\
\midrule
Embedding & 86.77 \\
Layer 0: Attention / FFN & 107.26 / 10.18 \\
Layer 1–31: Attention / FFN & 0.44 / 0.75 \\
Final RMSNorm & 2.51 \\
Logits Projection & 0.66 \\
\bottomrule
\end{tabular}
\end{table}


\subsection{Latency and Throughput Metrics} \label{se:Metrics}

We evaluate inference performance using six latency and throughput metrics. \textbf{Time to First Token (TTFT)} measures the initial delay before the first output token is produced, capturing prompt processing and setup overhead. \textbf{Time Between Tokens (TBT)} reflects the average interval between generating successive tokens and serves as an indicator of streaming efficiency. \textbf{Tokens per second} quantifies the overall generation speed and is used as the primary throughput metric. \textbf{Average latency per token} combines TTFT and TBT to express the total latency incurred for each generated token. \textbf{Batch throughput} captures the number of inference requests processed per second, relevant for multi-user or high-load scenarios. Finally, \textbf{total inference time} represents the complete wall-clock duration required to complete the end-to-end generation task, enabling assessment of overall responsiveness and overhead. Figure~\ref{fig:latency_throughput_comparison} compares these metrics across sequence lengths (100–400 tokens). SecureInfer consistently outperforms TEE-only and achieves near-GPU-level latency. We further break down prompt-phase latency in Table~\ref{tab:impact_parameters}, showing that embedding and Layer 0 attention/FFN dominate runtime (Table~\ref{tab:latency_breakdown}). These findings inform SecureInfer’s partitioning strategy.


\subsection{Runtime and Energy Efficiency}
To further dissect performance, we decompose the total inference time as

\begin{equation}
T_{\text{Infer}} = T_{\text{GPU}} + T_{\text{SGX}} + T_{\text{CPU}} + T_{\text{overhead}}
\end{equation}

\noindent where $T_{\text{GPU}}$ denotes the GPU execution time, $T_{\text{SGX}}$ the TEE enclave time, $T_{\text{CPU}}$ the CPU execution time, and $T_{\text{overhead}}$ the transition overhead incurred during transfers across GPU~$\leftrightarrow$~CPU~$\leftrightarrow$~SGX.

\begin{figure}[t]
    \centering
    \includegraphics[width=\linewidth]{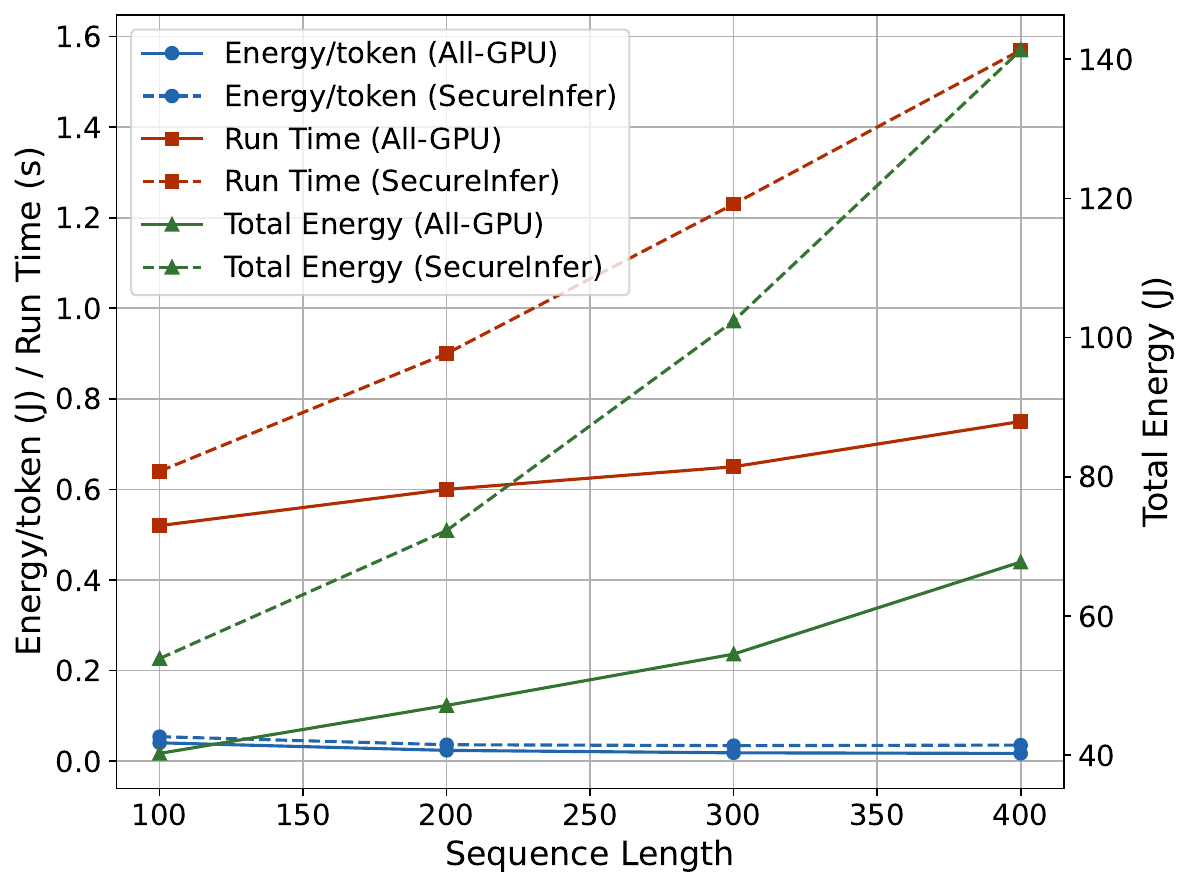}
    \caption{Energy per token, total energy, and run time versus sequence length and deployment mode.}
    \label{fig:seq-efficiency}
\end{figure}

SecureInfer reduces total latency by \textbf{4.7× over TEE-only} and retains only \textbf{2.06× overhead} versus GPU-only (Table~\ref{tab:latency-comparison}). Its throughput reaches 8.44 tokens/sec, offering a practical balance between performance and confidentiality.
Figure~\ref{fig:seq-efficiency} visualizes energy consumption trends. SecureInfer’s \textbf{energy per token} grows with sequence length and model depth, but remains within \textbf{2–3× of GPU-only}.

We also profiled data transfer latency between hardware boundaries. GPU-to-CPU transfers, often required for secure enclave operations, incur 0.2–0.3 ms per tensor, while enclave-to-GPU transfers are significantly cheaper ($<$0.01 ms). These findings suggest that caching or overlapping secure decryption with computation can further improve overall responsiveness.


\subsection{Energy and Efficiency Analysis} 
\label{se:energy}


To evaluate \textit{SecureInfer}'s feasibility on edge devices, we profile energy usage across transformer depth, batch size, and sequence length. Measurements are averaged over 10 runs using psutil, pynvml, and Intel PCM~\cite{psutil, Pynvml, Dementiev2025intel}. We report both total and per-token energy metrics to capture how overhead scales under different workloads.

\textbf{Depth Scaling.} 
Energy demand increases non-linearly with transformer depth. Table~\ref{tab:depth-effect} shows SecureInfer incurs 2–6× overhead depending on depth, due to enclave execution. Energy demand increases non-linearly with transformer depth. 

\textbf{Batch Size and Sequence Length.} 
Larger batch sizes amortize overhead and improve energy efficiency. Table~\ref{tab:batch-scaling} confirms that increasing batch size from 1 to 8 reduces energy/token and runtime. As shown in Table~\ref{tab:power-vs-seq}, SecureInfer incurs only modest increases in average power demand across varying sequence lengths, with overhead dropping from 1.10× to 0.99× as sequence length grows from 100 to 400 tokens. This trend suggests that SecureInfer's additional memory and computation costs are effectively amortized over longer inputs, reinforcing its suitability for real-time workloads that involve moderate to long prompts."

\textbf{Layer-wise Profiling.}
Table~\ref{tab:energy-summary} shows linear layers dominate GPU energy (4–5 J), while LoRA and nonlinear ops are lightweight ($<$0.3 J). This supports SecureInfer’s design to isolate security-critical but lightweight components and offload costly operations.

\begin{table}[h]
\centering
\caption{Impact of model depth on energy per token.}
\begin{tabular}{cccc}
\toprule
Layers & all\_gpu & SecureInfer & Overhead \\
\midrule
1      & 0.0076   & 0.0353  & 4.6$\times$ \\
2      & 0.0332   & 0.0636  & 1.9$\times$ \\
2      & 0.0527   & 0.0949  & 1.8$\times$ \\
4      & 0.0718   & 0.1305  & 1.8$\times$ \\
\bottomrule
\end{tabular}
\label{tab:depth-effect}
\end{table}

\begin{table}[h]
\centering
\caption{Impact of sequence length on average power demand (W).}
\begin{tabular}{cccc}
\toprule
Seq Len & all\_gpu (W) & SecureInfer (W) & Overhead \\
\midrule
100     & 84.4         & 93.03       & 1.10$\times$ \\
200     & 87.2         & 89.11       & 1.02$\times$ \\
300     & 93.24        & 92.76       & 0.99$\times$ \\
400     & 101.0        & 100.18      & 0.99$\times$ \\
\bottomrule
\end{tabular}
\label{tab:power-vs-seq}
\end{table}

\begin{table}[h]
\centering
\caption{Total energy (J) for increasing batch size (transformer layers = 1).}
\begin{tabular}{cccc}
\toprule
Batch & all\_gpu (J) & SecureInfer (J) & Overhead \\
\midrule
1     & 60.88   & 141.3  & 2.3$\times$ \\
2     & 62.68   & 134.74  & 2.2$\times$ \\
4     & 63.33   & 133.47  & 2.1$\times$ \\
8     & 59.74   & 133.72  & 2.2$\times$ \\
\bottomrule
\end{tabular}
\label{tab:batch-scaling}
\end{table}

\begin{table}[h]
\centering
\caption{Representative GPU Energy Consumption per Layer Type}
\label{tab:energy-summary}
\begin{tabular}{lcc}
\toprule
\textbf{Layer Type} & \textbf{Duration (s)} & \textbf{Energy (J)} \\
\midrule
Linear (e.g., w1, w2, wo) & 0.10 -- 0.25 & 4 -- 5 \\
Nonlinear (silu, norm) & $<$ 0.05 & $<$ 0.05 \\
LoRA Adapter & $\approx$ 0.11 & 0.1 -- 0.3 \\
\bottomrule
\end{tabular}
\end{table}

\begin{table}[!t]
\centering
\caption{Performance Comparison Across Deployment Modes (Seq\_Len=300)}
\label{tab:latency-comparison}
\resizebox{\columnwidth}{!}{%
\begin{tabular}{lcccc}
 \toprule
\textbf{Metric} & \textbf{GPU Only} & \textbf{CPU Only} & \textbf{TEE Only} & \textbf{SecureInfer} \\
\midrule
Time to First Token (ms) & 57.41 & 469.13 & 563.64 & 122.70 \\
Time Between Tokens (ms) & 57.46 & 447.29 & 553.30 & 118.37 \\
Tokens/sec & 17.40 & 2.23 & 1.81 & 8.44 \\
Avg. Latency per Token (ms) & 57.46 & 447.72 & 553.51 & 118.45 \\
Batch Throughput (req/sec) & 0.35 & 0.04 & 0.04 & 0.17 \\
Total Time (s) & 2.87 & 22.39 & 27.68 & 5.92 \\
Tokens Generated & 50 & 50 & 50 & 50 \\
\bottomrule
\end{tabular}%
} 
\end{table}

\begin{figure}[h]
    \centering
    \includegraphics[width=0.9\linewidth]{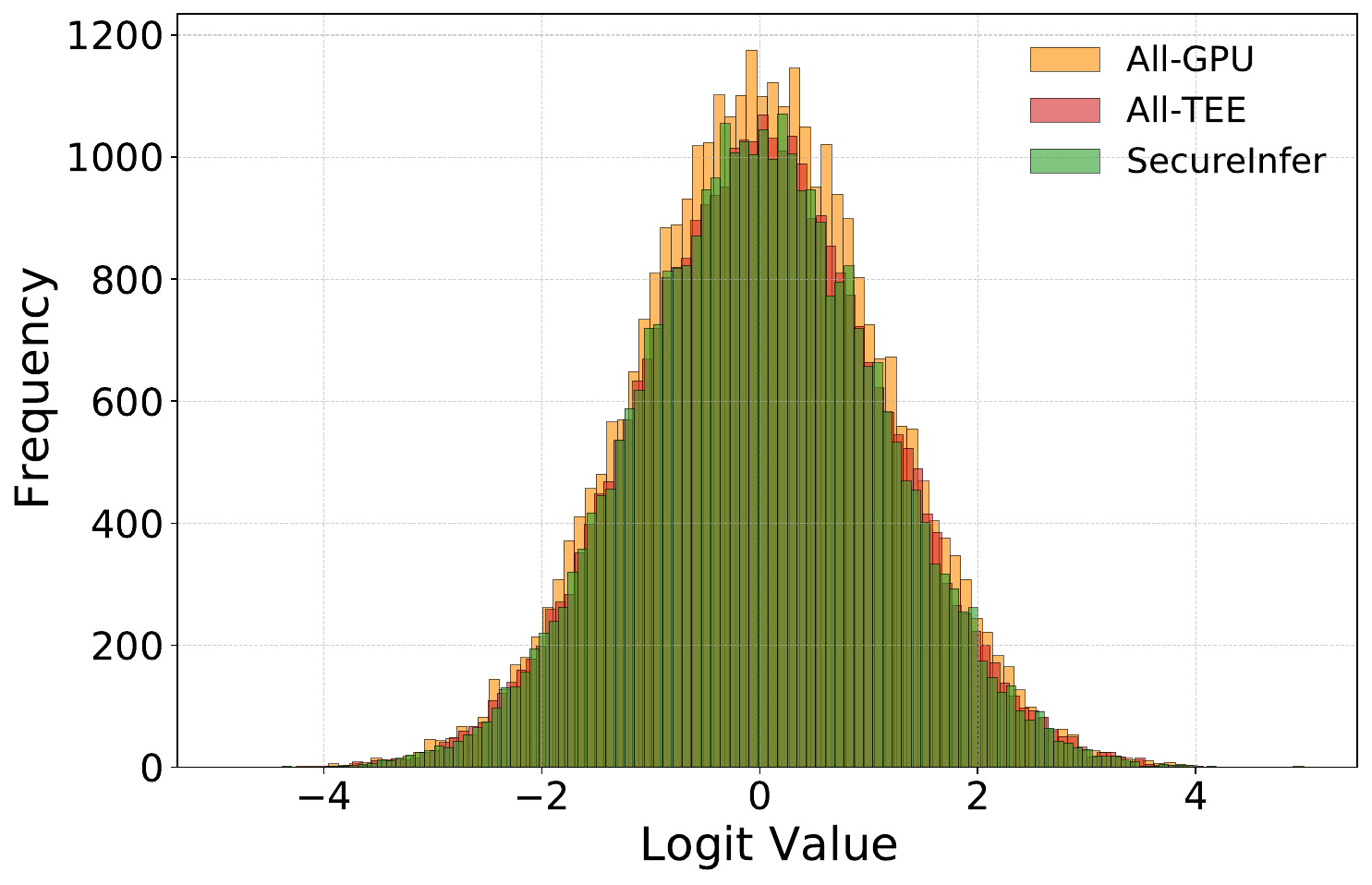}
    \caption{Logit distribution across different deployment modes}
    \label{fig:logit-distribution}
\end{figure}


\begin{table}[t]
\centering
\scriptsize
\setlength{\tabcolsep}{2.5pt}
\begin{threeparttable}
\caption{Extraction results on decoder-only models (Gen. task, 1000 queries).}
\label{tab:gen_extraction}
\begin{tabular}{@{}l l l c c c c c c@{}}
\toprule
\textbf{Dataset} & \textbf{Victim} & \textbf{Student} & \textbf{BLEU}$_B$ & \textbf{BLEU}$_S$ & \textbf{R-F1} & \textbf{F2-F1} & \textbf{R-L} & \textbf{Tok. Match} \\
\midrule
SQuAD   & GPT-2   & DistilGPT2  & 0.1310 & 0.6289 & 0.7889 & 0.7985 & 0.8213 & 0.9418 \\
PIQA    & LLaMA2   & TinyLlama   & 0.1257 & 0.4532 & 0.6072 & 0.5577 & 0.6584 & 0.8258 \\
\bottomrule
\end{tabular}
\begin{tablenotes}
\item \textbf{Abbreviations:} BLEU$_B$ = Baseline, BLEU$_S$ = Student, R-F1 = ROUGE-F1, R-L = ROUGE-L, Tok. Match = Token Match.
\end{tablenotes}
\end{threeparttable}
\end{table}

\begin{table}[t]
\centering
\scriptsize
\setlength{\tabcolsep}{2.5pt}
\caption{Extraction results on classification models.}
\label{tab:classification_extraction_simple}
\begin{tabular}{@{}l l c c@{}}
\toprule
\textbf{Victim Model} & \textbf{Surrogate Model} & \textbf{Agreement Rate} ↑ & \textbf{KL Div.} ↓ \\
\midrule
BERT-340M & DistilBERT-66M    & 0.940 & 0.080 \\
BERT-340M & BERT-Small-110M   & 0.990 & 0.068 \\
\bottomrule
\end{tabular}
\end{table}


\begin{table}[h]
\centering
\small
\setlength{\tabcolsep}{4pt}
\caption{Defense Evaluation for LLaMA2; Same model setting as attack (Table~\ref{tab:gen_extraction}), but with defense.}
\label{tab:llama2-defense-compact}
\resizebox{0.9\columnwidth}{!}{%
\begin{tabular}{lcccccc}
\toprule
& \textbf{B-1} & \textbf{B-2} & \textbf{B-3} & \textbf{B-4} & \textbf{Tok Match} & \textbf{Cont Div} \\
\midrule
\textbf{Score} & 0.1098 & 0.1018 & 0.1126 & 0.1180 & 0.5643 & 0.2102 \\
\bottomrule
\end{tabular}}
\end{table}



\subsection{Defense Against Model Extraction}

To evaluate defense effectiveness, we implement a black-box model extraction attack where an adversary queries the target model and reconstructs a surrogate. Following prior work~\cite{Knockoff, tramer2016stealing, ML-Doctor}, we measure success using agreement rate (classification) and BLEU/ROUGE scores (generation).

Table~\ref{tab:classification_extraction_simple} and Table~\ref{tab:gen_extraction} report results when no defenses are applied. Surrogate models achieve high agreement and BLEU scores, indicating successful extraction. SecureInfer applies two complementary defenses: \textbf{(i) TEE-Based Partitioning}: sensitive components (e.g., FFNs, attention projections) are kept inside SGX to prevent memory snooping; and \textbf{(ii) Logits Perturbation}: API outputs are lightly obfuscated using response randomization. These defenses significantly reduce attack effectiveness, lowering BLEU scores below 0.12 and limiting token match to 56.4\%, as shown in Table~\ref{tab:llama2-defense-compact}.

\subsection{Output Consistency and Fidelity}  \label{se:Model-Output} 
We examine whether SecureInfer affects model behavior by comparing output logits across modes. Figure~\ref{fig:logit-distribution} shows that logit distributions from SecureInfer closely match those from GPU-only and TEE-only executions. Minor shifts are attributed to floating-point precision differences, confirming fidelity preservation.

\section{Discussion}

SecureInfer introduces a practical trade-off between security and performance, but faces several challenges. Our current implementation faces challenges due to the tight coupling of transformer blocks, requiring careful balancing of security and performance. Additionally, hardware dependency, particularly concerning enclave memory constraints, imposes operational limits on input size and complexity.


Nevertheless, our defensive techniques generalize well and can extend protection to broader attack classes, including fine-tuning-based model repurposing. Future enhancements include dynamic runtime partitioning and compiler-level optimizations using frameworks such as TVM~\cite{Chen2025apache}, vLLM~\cite{kwon2023efficient}, and MLC-LLM~\cite{mlc-llm}.

In addition, integrating advanced quantization strategies, such as mixed-precision arithmetic with obfuscated rounding, and supporting continuous batching schemes may further reduce trusted computing overhead without compromising throughput. These extensions would strengthen the practicality of across diverse deployment settings.

Overall, \textit{SecureInfer} marks a step toward a secure-by-design LLM paradigm.

\section{Conclusion}



We presented \textbf{SecureInfer}, a hybrid TEE–GPU framework for secure and efficient LLM inference on edge devices. SecureInfer isolates privacy-critical components, such as attention projections and adaptation layers within an SGX enclave while securely offloading high-throughput computations to the GPU. To further reduce information leakage, it applies lightweight perturbations to output logits.

Our evaluation demonstrates that \textbf{SecureInfer} offers strong resilience against model extraction attacks, while achieving a \textbf{3.7$\times$ throughput gain} over \textit{TEE-only} deployments. Compared to the \textit{GPU-only} baseline, it incurs a modest \textbf{2.06$\times$} latency overhead with minimal degradation in output fidelity. These results position SecureInfer as a practical and scalable solution for securely deploying LLMs under real-world resource and privacy constraints.

\section*{Acknowledgement}
We thank the anonymous reviewers for their helpful feedback and time. This work was partially supported by the US National Science Foundation (Awards: 2419843). The views expressed are those of the authors only, not of the funding agencies.

\bibliographystyle{IEEEtran}
\bibliography{references}

\end{document}